\documentclass[twocolumn,english,journal]{IEEEtran}

\ifCLASSINFOpdf
\else
\fi

\usepackage[T1]{fontenc}
\usepackage{babel}
\usepackage{array}
\usepackage{calc}
\usepackage{amsmath}
\usepackage{amsthm}
\usepackage{amssymb}
\usepackage{graphicx}
\usepackage{cite}
\usepackage{algorithm}
\usepackage{subfigure}
\usepackage{color}

\usepackage{enumitem}
\usepackage{multirow}
\usepackage{hhline}

\usepackage[unicode=true,
bookmarks=true,bookmarksnumbered=true,bookmarksopen=true,bookmarksopenlevel=1,
breaklinks=false,pdfborder={0 0 0},backref=false,colorlinks=false]
{hyperref}
\hypersetup{pdftitle={Your Title},
	pdfauthor={Your Name},
	pdfpagelayout=OneColumn, pdfnewwindow=true, pdfstartview=XYZ, plainpages=false}

\makeatletter

\let\oldforeign@language\foreign@language
\DeclareRobustCommand{\foreign@language}[1]{%
	\lowercase{\oldforeign@language{#1}}}

\makeatother

\allowdisplaybreaks

\makeatletter
\newcommand{\thickhline}{%
	\noalign {\ifnum 0=`}\fi \hrule height 1.5pt
	\futurelet \reserved@a \@xhline
}
\newcolumntype{"}{@{\hskip\tabcolsep\vrule width 1pt\hskip\tabcolsep}}
\makeatother

\newcolumntype{P}[1]{>{\centering\arraybackslash}p{#1}}
\newcolumntype{M}[1]{>{\centering\arraybackslash}m{#1}}

\begin{document}

	\title{A Verifiable Framework for Cyber-Physical Attacks and Countermeasures in a Resilient Electric Power Grid}

	\author{\IEEEauthorblockN{Zhigang Chu$^1$, Andrea Pinceti$^1$, Ramin Kaviani$^1$, Roozbeh Khodadadeh$^1$, Xingpeng Li$^1$, Jiazi Zhang$^1$, Karthik Saikumar$^1$, Mostafa Sahraei-Ardakani$^1$, Christopher Mosier$^2$, Robin Podmore$^3$, Kory Hedman$^1$, Oliver Kosut$^1$, and Lalitha Sankar$^1$\\}
		\IEEEauthorblockA{1. Arizona State University \hspace{2.5cm}
			2. PowerData \hspace{2.5cm}
			3. IncSys} \hspace{2.5cm}
	\thanks{Manuscript received: XX XX, XXXX; accepted: XX XX, XXXX. Date of CrossCheck: XX XX, XXXX. Date of online publication: XX XX, XXXX. 
	
	This material is based upon work supported by the National Science Foundation under Grants No. CNS-1449080 and OAC-1934766 and the Power System Engineering Research Center (PSERC) under projects S-72. 
	
	$^1$ The authors are with the School of Electrical, Computer and Energy Engineering at Arizona State University, Tempe, AZ, USA.
	
	$^2$ The author is with PowerData Corporation, Redmond, WA, USA.
	
	$^3$ The author is with Incremental Systems Corporation, Issaquah, WA, USA.
	}
	}
    
	\maketitle

	\begin{abstract}
	In this paper, we investigate the feasibility and physical consequences of cyber attacks against energy management systems (EMS). Within this framework, we have designed a complete simulation platform to emulate realistic EMS operations: it includes state estimation (SE), real-time contingency analysis (RTCA), and security constrained economic dispatch (SCED). This software platform allowed us to achieve two main objectives: 1) to study the cyber vulnerabilities of an EMS and understand their consequences on the system, and 2) to formulate and implement countermeasures against cyber-attacks exploiting these vulnerabilities. Our results show that the false data injection attacks against state estimation described in the literature do not easily cause base-case overflows because of the conservatism introduced by RTCA. For a successful attack, a more sophisticated model that includes all of the EMS blocks is needed; even in this scenario, only post-contingency violations can be achieved. Nonetheless, we propose several countermeasures that can detect changes due to cyber-attacks and limit their impact on the system. 

	\end{abstract}
	

	\IEEEpeerreviewmaketitle
	\global\long\def\figurename{Fig.}
	\global\long\def\tablename{TABLE}
	
	\section{Introduction}
	
	The last decade has seen cyber-attacks become a new source of potential failure in power systems. Several cyber-attack incidents have occured on real-world systems, including the Stuxnet malware targeting the Iranian power grid \cite{StuxnetIran}, and the cyber-attacks that caused blackouts in Ukraine \cite{UkraineAttack,Ukraine2017a}. Reportedly, the U.S. power grid is under almost continuous attack \cite{ReportContinuousAttacks}, and there is evidence that attackers have successfully hacked into U.S. power systems before \cite{USAttack2018,USAttack2018_2}.
	
	This paper focuses on false data injection (FDI) attacks that involve a malicious attacker replacing a subset of measurements with counterfeits. FDI attacks can be designed to target system states \cite{Liu2009,Kosut2011,Hug2012}, system topology \cite{Jzhang2016,Liu2017}, generator dynamics \cite{Kundur2010}, and energy markets \cite{Xie2011}. Consequences of FDI attacks are often evaluated via optimization problems, \textit{e.g.,} FDI attacks that aim to: maximize line power flow \cite{Liang2015}, maximize operating cost \cite{Yuan11,Yuan2012}, or change locational marginal prices \cite{Jia2014}. However, the majority of work in the literature considers consequences only via unrealistically simple assumptions about  system operations, \textit{e.g.,} DC state estimation (SE) and DC optimal power flow (OPF). Modern energy management systems (EMSs) typically involve more complicated components, including real-time contingency analysis (RTCA) and security constrained economic dispatch (SCED), to ensure $N-1$ reliable operation. In addition to the constraints modeled in DCOPF, SCED determines the system dispatch by taking into account security constraints reported by RTCA, as well as constraints on ramp rates and reserves. To better understand the attack consequences on state-of-the-art EMSs, and to design countermeasures, a Java-based EMS simulation platform (consisting of a physical system simulator, state estimator, RTCA, and SCED) was developed. 
	
    This simulation platform enables us to answer two fundamental questions. Given the documented weaknesses of some systems (e.g. SE) and the increase in number and types of cyber threats targeted at the energy sector, our first goal is to investigate 
	further vulnerabilities which might be present in state-of-the-art EMSs. The second main objective of this study is to understand the practical consequences of cyber-attacks that exploit these vulnerabilities. In a more general sense, addressing these questions allows us to determine if the added complexity in control and monitoring systems provides resiliency against cyber-attacks, or if the approximations made in modeling these systems introduce new vulnerabilities. This understanding represents the foundation for new and improved algorithms and countermeasures to create a more resilient electric power grid.
	
	The key contributions of this work and the lessons learned through this research are as follows:
	
	\begin{itemize}
		\item Creating a Java-based EMS simulation platform that includes a physical system simulator, state estimator, RTCA, and SCED to mimic the actual operations of state-of-the-art power systems.
		\item Investigating vulnerabilities that are present in state-of-the-art EMSs through designing FDI attacks and testing their physical consequences using the simulation platform. This work shows that, by satisfying the N-1 reliability requirements, the base case solution is N-1 secure and, thus, it is hard for attackers to cause a pre-contingency overload. 
		\item Developing several countermeasures against such attacks by exploiting trusted historical data and fundamental knowledge of power systems. Extensive testing shows that these countermeasures can efficiently detect FDI attacks. 
	\end{itemize}

    The remainder of the paper is organized as follows. In Section \ref{sec:EMS} the core functionalities of the EMS simulation platform are described. In Section~\ref{sec:EMSimplementation} the software implementation of the platform and its graphical user interface are discussed. Section~\ref{sec:Attack} presents an overview of load redistribution attacks and their impact on real-world systems operations. Finally, in Section~\ref{sec:Countermeasure} several countermeasures against these attacks are introduced and the machine learning-based detection algorithm that has been implemented within the EMS platform is described. Supporting documents, including an Appendix, source code, and additional figures, can be found at \cite{Pincetigithub}.
	
	
	\section{\label{sec:EMS}Energy Management System Emulator Platform: Core Functionalities}
\begin{figure*}
		\centering
		\includegraphics[trim=0 0.3cm 0 0.3cm, scale=0.7]{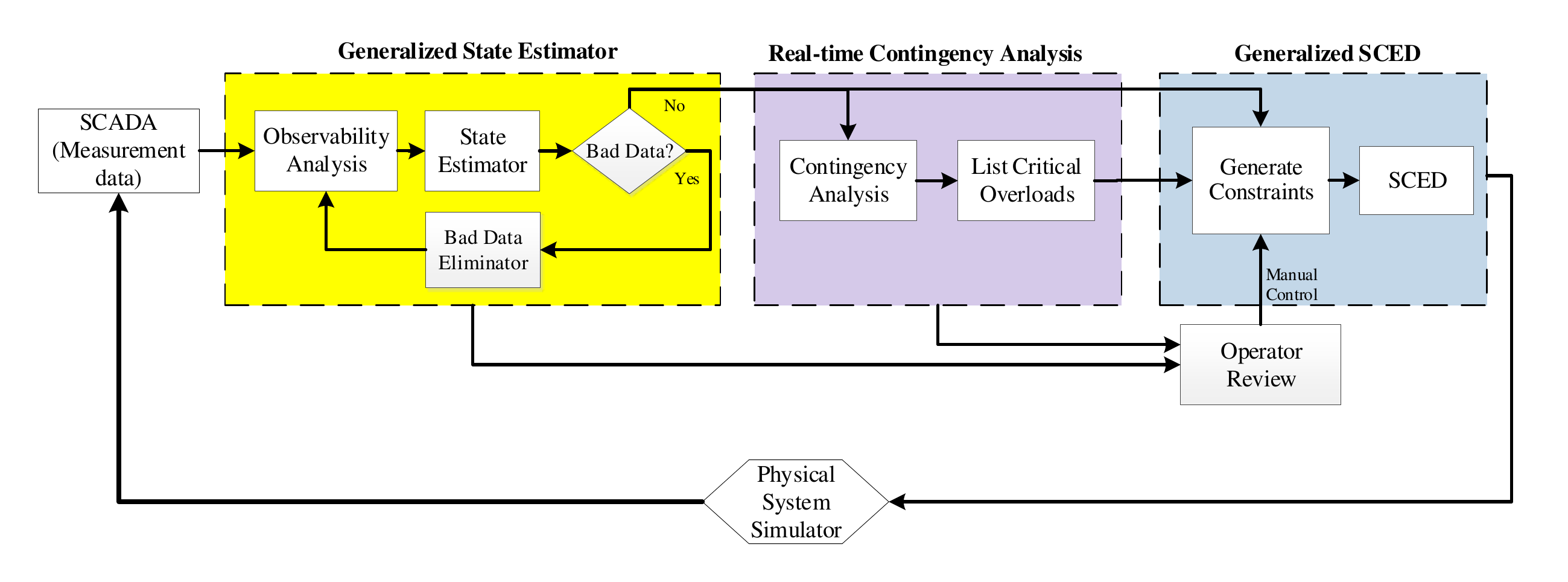}\caption{EMS simulation platform}\label{fig:EMS}
	\end{figure*}

	In this section, the Java-based EMS emulator platform, which we developed, is described. In the context of power system operations, an EMS is a critical tool that allows for the secure and efficient functioning of an electric grid. An EMS includes monitoring capabilities such as supervisory control and data acquisition (SCADA) systems, as well as automated optimization and control tools.  
	Figure \ref{fig:EMS} illustrates the operation of our EMS platform. The physical system simulator utilizes a fast-decoupled power flow algorithm to model the behavior of physical power systems. The SE uses the measurements collected by SCADA to estimate the system operating states and remove bad data. RTCA is then performed based on the results of SE to obtain a list of the existing and potential critical contingencies. These contingencies are flagged, passed to SCED, and then the SCED module generates the corresponding security criteria to protect against the critical contingencies. The SCED solves a linear program to find the most economic generation dispatch that ensures reliable operation. The backbone of the EMS emulator is built upon commercial products developed by IncSys \cite{IncSys}, PowerData \cite{OpenPA}, and work developed at ASU for an ARPA-E GENI project, an ARPA-E NODES project, and this NSF/DHS funded project. 
    
    \subsection{\label{sec:PF}Physical System Simulator}
\begin{figure}
    	\centering
    	\includegraphics[trim = 100mm 68mm 9mm 50mm, clip, width=14.2cm]{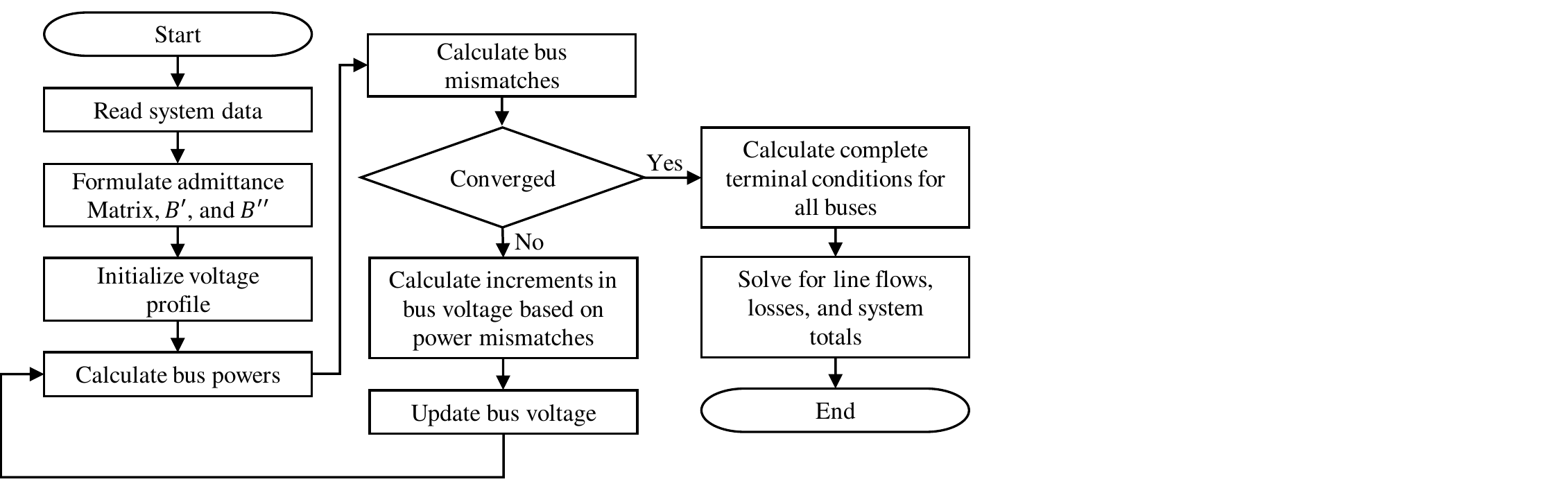}
    	\caption{Flowchart of the Power Flow Algorithm}
    	\label{pf-chart}
    \end{figure}

    The physical system simulator uses the fast decoupled Newton-Raphson (FDNR) method to solve an AC power flow based on the initial generation dispatch. This method is a simplified version of the Newton-Raphson method, considering several assumptions which are discussed in \cite{Load1974Stott}, thoroughly.
    The FDNR method considerably reduces the computation at each iteration and the time of convergence since using the simplified admittance matrices $B'$ and $B''$ curtails three-quarter of the full Newton power flow Jacobian matrix. Moreover, $B'$ and $B''$ are constant, which means they only need to be computed once at the beginning of the algorithm, except for changes related to generation volt-ampere reactive (VAR) limiting in $B''$ matrix, which needs to be updated. The general flow of the algorithm is shown in Fig. \ref{pf-chart}.
    
     In \cite{podmore1972digital}, the authors describe the implementation of LinkNet, a structure for the computer representation of networks which is at the base of OpenPA. They also propose algorithms to count connected islands in the graph, find a certain bus, branch, or connection list of a bus, and calculate and triangulate the Jacobian matrix during the fast decoupled power flow analysis. These algorithms are implemented in our physical system simulator. The OpenPA core, which has been written using LinkNet, allows the user to efficiently compute the triangular network matrices, build bus impedance matrices, and run nodal iterative power flow. 
    
    Several adjustments and improvements were added to this general algorithm to allow the user to configure the convergence threshold and, if the mismatches are close to this threshold, distributing the slack mismatch among generators proportionally to their maximum active power output. It also does not allow generators to exceed their capability limits during this process. In addition, our physical system simulator detects if there are multiple energized islands in the system and runs the power flow for each island separately and reports the results. During this process, reference buses to each island are automatically assigned. These reference buses are chosen based on node degree of the bus and its generation capability. Static Var Compensator (SVC) devices are also monitored and applies appropriate changes to the $B''$ matrix as the algorithm progresses.

	\subsection{\label{sec:SE}State Estimation}
	State estimation enables real-time monitoring of power systems. Measurements, including branch real and reactive power flows, bus real and reactive power injections, and bus voltage magnitudes, are collected by SCADA and sent to the system control center to estimate the system states, namely the bus voltage magnitudes and phase angles. The solution of state estimation provides a starting point for the following EMS processes including real-time contingency analysis, voltage security assessment tools, voltage support service, and other advanced network applications \cite{AburBook}. 
	
    Our generalized SE includes observability analysis (OA), bad data detector (BDD), and bad data eliminator (BDE) as shown in Fig. \ref{fig:EMS}. OA determines whether all system states can be estimated from available measurements; if successful, the system is observable and SE is performed on the whole system. Otherwise, OA will output several observable islands, and SE is performed on each island. Our OA algorithm is adopted from \cite{Clements1983} and its detailed description is given in Appendix A \cite{Pincetigithub}. The system states $\hat{x}$ are estimated using Weighted Least-Squares (WLS) method from measurements $z=h(x)+e$ as described in \cite{AburBook}, where $z$ is the measurement vector, $x$ is the vector of true states, $h(\cdot)$ is the non-linear relationship between measurements and states, and $e$ is additive noise. Givens rotation \cite{Givens} is used to perform orthogonal factorization when solving the normal equation. The detailed SE algorithm is given in Appendix B \cite{Pincetigithub}. The estimated states $\hat{x}$ are then used to recalculate the measurements $\hat{z}=h(\hat{x})$, and the BDD applies a $\chi^2$-test to detect bad data. If bad data exists, the measurement that has the largest normalized residual is eliminated by the BDE. OA and SE are then performed on the remaining measurements until the $\chi^2$-test is passed. 
	
	\subsection{\label{sec:RTCA}Real-time Contingency Analysis}
	After SE determines the power system operating states, RTCA is run to evaluate the impact of power system component failures. Industry planning and operating criteria often enforce $N-1$ reliability, which requires that a system must operate in a stable and secure manner following any single transmission or generation outage. This is ensured by taking each system element (generating unit, transmission line, transformer) out of service one at a time, and re-running power flow to see if there are any (potential) violations to voltage and line limits. 
	
\begin{figure}
		\includegraphics[trim = 79mm 88mm 1mm 48mm, clip, width=16cm]{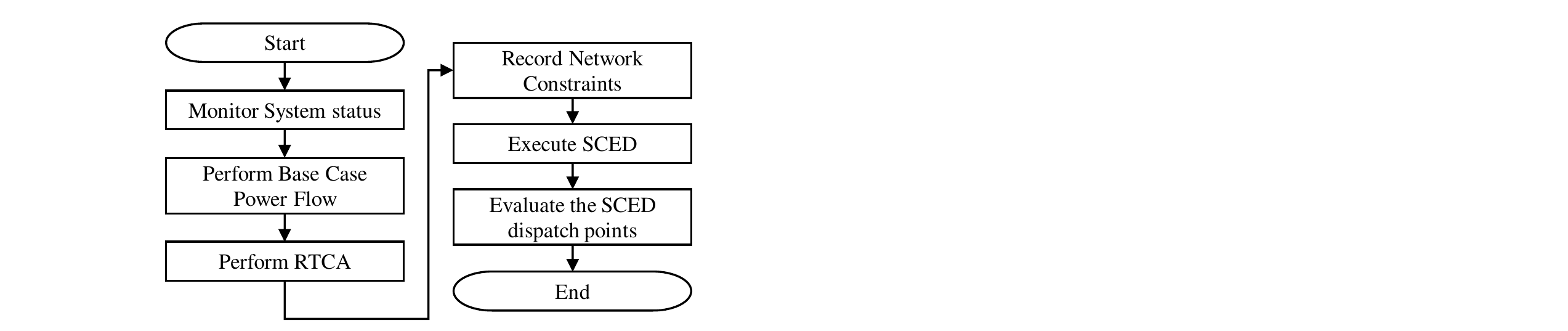}
		\caption{Flowchart of RTCA and SCED. Source: \cite{xingpeng2019}}
		\label{rtca-sced}
	\end{figure}

	Our RTCA focuses on branch (transmission line and transformer) outages and violations. Outages of radial branches can break the system into islands and are out of the scope of this paper. All the radial branches are identified in the system and they are excluded from the contingency list. Then, the contingency rating of every branch is determined, as branches are typically allowed to safely carry a relatively higher power flow for a short period of time compared to their long-term power ratings. Subsequently, AC power flow is run as described in Sec. \ref{sec:PF} by taking one branch out of service, sequentially from the contingency list, to see if it leads to violations or potential violations on any other branches. If any of the other branches have flows higher than their contingency ratings (violations) or close to them (warnings), their power flow magnitudes are recorded, and this branch outage is flagged as a critical contingency. This process is performed until outages of all branches in the contingency list are processed. Finally, RTCA creates a list of critical contingencies along with the corresponding violations, branches with warnings, and their power flow magnitudes. This information is sent to the subsequent SCED function as transmission security constraints to determine the generation dispatch for the next operating period; this process is illustrated in Fig.~\ref{rtca-sced} (figure courtesy of Dr. Xingpeng Li, \cite{xingpeng2019}).
	
	\subsection{\label{sec:SCED}Security-constrained Economic Dispatch}
	In today's power system operation, SCED is one of the key functions of the EMS. It is an optimization program that finds the least cost generation satisfying all the operation and security constraints. Though the AC power flow model is more accurate, it is not used due to its computational complexity. Typically, SCED is a linear program (LP) since all generators' status are considered as fixed input and DC power flow model is used. Thus, DC power flow model based SCED is simple and optimality can be guaranteed.
	
	However, the optimal generation dispatch point output by the SCED needs to be applied on the physical AC system. Therefore, properly approximating the AC system in a DC SCED model is crucial to the SCED formulation. The three following types of AC approximation techniques are used to better mimic the AC system behavior in order to provide a better SCED solution. A more in-depth and detailed description of the following considerations can be found in \cite{xingpeng2019}. 
	
	(i) The first approximation technique is to model AC system losses in the DC SCED model. Without handling the losses, the SCED  dispatch solution is not sufficient to cover the system loads and real power losses, and could potentially lead to AC power flow non-convergence issues. Five different options have been implemented to handle the losses; the first two options are to simply distribute the real loss on load or generator buses respectively, and in the three remaining options losses can be added on each branch to the receiving-bus, sending-bus, or half of the loss to the receiving-bus and half to the sending-bus as virtual loads, respectively.
	
	(ii) The second AC approximation technique is to derate the normal and emergency branch MVA ratings to obtain the branch MW ratings, by taking the reactive power flows into consideration. If the SCED uses AC ratings as branch limit constraints, the resulting generation dispatch from SCED may create real power flows on lines at their apparent power limits. Along with the reactive power flows, it may result in branch overflows in terms of apparent power. Thus, the branches can be derated to avoid this issue. Given a branch $k$ with AC branch limit $S_{k,\text{max}}$, the limit of this branch in base case can be modeled as
	\begin{equation}
		P_{k,\text{max}}=\sqrt{S_{k,\text{max}}^2-\text{max}(Q_{kf}^2,Q_{kt}^2)}, \label{eq:Pk_max}
	\end{equation}
	where $Q_{kf}$ and $Q_{kt}$ are the \textit{from} and \textit{to} end branch reactive power flows, respectively. In contingency case, $S_{kc,\text{max}}$, the emergency MVA rating of branch $k$, is used instead of $S_{k,\text{max}}$ to determine the contingency case branch limit $P_{kc,\text{max}}$.
	
	(iii) Finally, the last approximation is to calculate the branch real power flows (both base case and contingency case) by summing up their pre-SCED AC values and the DC approximated changes caused by the SCED re-dispatch, so that they are not computed using just purely DC power flow equations. 
	
	The complete SCED formulation used in our platform can be found in Appendix C \cite{Pincetigithub}; here, the main characteristics and capabilities of the algorithm we implemented are summarized. The SCED aims to minimize the sum of operating cost and reserve cost, assuming a piece-wise linear cost function for all generators. Consistent with industry (ISO) practice, some relaxations to SCED are applied: there are load shedding variables in both base case and contingency case, as well as a slack variable for violation of the generator minimum input. Penalty factors are given to these slack variables in the objective function, so that under normal operations, all slack variables should be zero. Non-zero load shedding indicates that the system is unable to serve the load given the transmission capacity and generator ramp rate limits, while non-zero generator lower bound slack variables indicates that there are too many committed units. The constraints in our SCED include (both in base case and contingency case) power balance equations, load equations, branch flow limits (a power transfer distribution factor (PTDF) based formulation is adopted in calculating branch power flows from bus power injections), unit generation equations, ramp rate limits, and interface limits. An interface limit constrains the power flow from one area to another to avoid voltage issues, generally through several major transmission lines. The interface limit is typically lower than the algebraic sum of the thermal limits of all the lines constituting the interface, but it can be included in SCED based on the operators' prior knowledge. In addition, generation limits and spinning reserve limits are also included. Though generator contingencies are not modeled, the SCED formulation contains a reserve requirement constraint, which ensures a system wide procurement of reserves that would be sufficient to cover the loss of an arbitrary generator.
	
	\section{Platform Implementation}\label{sec:EMSimplementation}
	\subsection{Web-based Graphical User Interface}
	To better study the process of FDI attacks and demonstrate their consequences, a web-based graphical user interface (GUI) is developed which allows the user to interact with all of the platform blocks. The environment is designed to mimic real world power system operations with an interface that aims to be easy to use and familiar to system operators. A user can view power system connectivity information and also run power flow, state estimation, contingency analysis, and SCED on a given system. With these capabilities, the platform can simulate the behavior of a power transmission system and its EMS in (almost) real-time.
	
	The platform has been created using a combination of Java and Scala programming languages. It uses OpenPA as a computational back-end to handle the power system model and run many of the power system calculations. The web-based GUI translates user commands to OpenPA functions and the results are returned back through the web application to a highly customizable visualization tool written using CytoScapeJs. The whole platform has been designed in a modular style with full decoupling between back-end and front-end.
	Moreover, both back-end and front-end have a modular structure to make extensions and improvements more straightforward. All the details on the software design of the platform and web application are described in \cite{roozbeh}. 
	
	\begin{figure*}
		\centering
		\includegraphics[trim=0 0.3cm 0 0.3cm, scale=0.55]{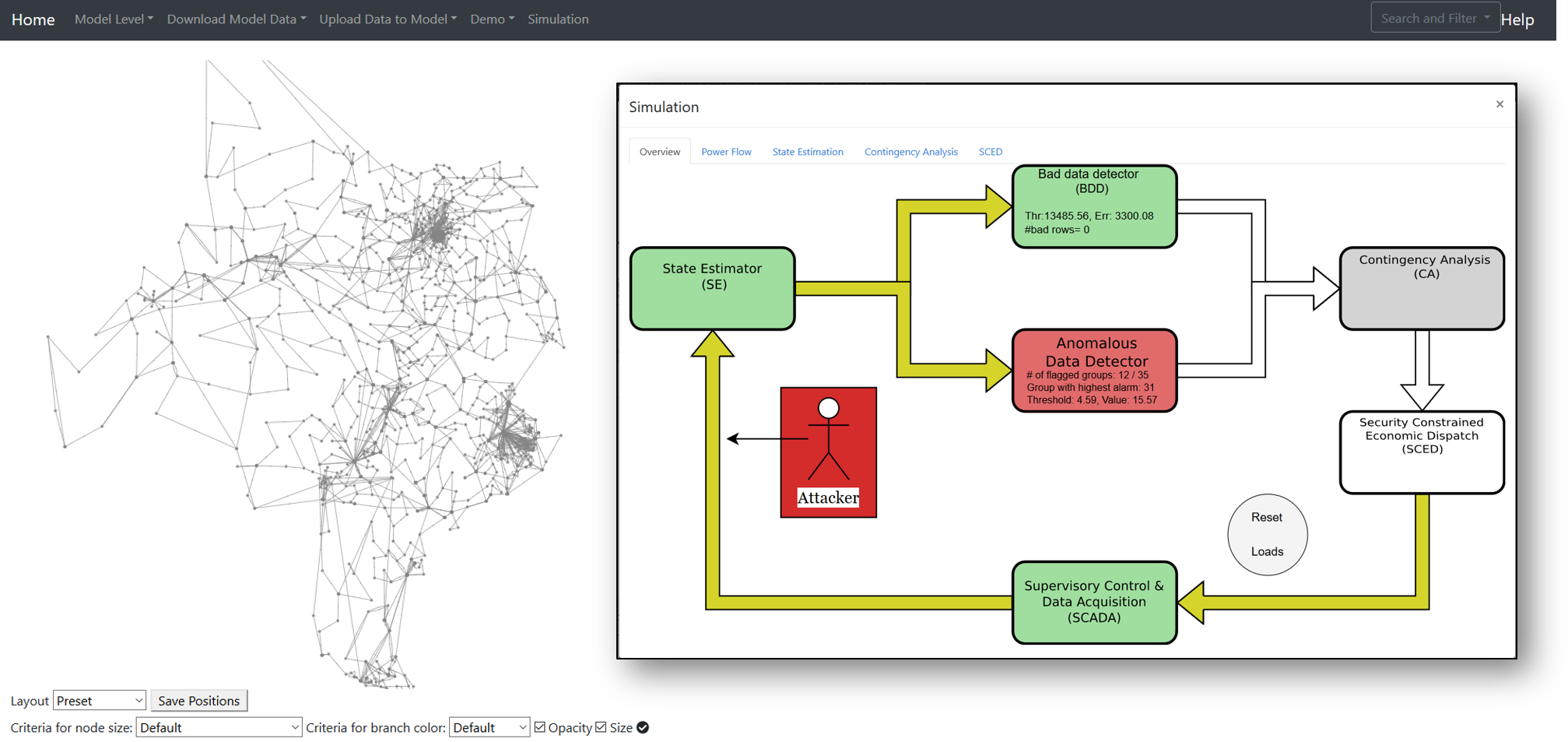}\caption{EMS Platform Graphical User Interface}\label{fig:GUI}
	\end{figure*}
	
	Figure~\ref{fig:GUI} shows a screenshot of the web-based GUI highlighting the main components of the platform. On the left is an interactive visualization of the system graph: the user can explore the system by zooming, panning, and obtaining details of each component by simply clicking on them (transmission line parameters, load and generator data, etc.). The panel on the right represents the core of the EMS platform; all of the functions can be set up and run via the GUI. In addition to the traditional EMS blocks, the diagram also includes the novel anomalous data detector which we have designed to detect FDI attacks against SE. The details of the detection algorithm are discussed in Section~\ref{sec:DetectionAlgo}. Figure~\ref{fig:GUI} illustrates that the attack detector can be used in parallel to the traditional bad data detector; as explained in the next section, an unobservable FDI attack will not trigger the residue-based detector. The screenshot of the interface shows that when the attack is injected and SE is performed over the malicious data, our detector is able to flag the anomaly before the false states are used to perform RTCA and SCED. 
	
	Finally, backed by OpenPA, the platform makes comprehensive and detailed reports available for download to the user. For example, the user can examine every step of the power flow solution or view contingencies or SCED solution reports in detail. Using these reports, informative diagrams can be created to give profound insight into the designed attacks.
	
	The code of the web-based GUI interface and the proposed attack detector as well as the power systems models used for our simulations are publicly available at \cite{Pincetigithub}. This GitHub repository also contains a video showcasing the EMS platform: it presents all of the main functions and it shows the complete process of a cyber-attack, from the injection of the falsa data to the physical consequences on the system. The video can also be found at: https://youtu.be/UDAMI3gi3WI.

	\subsection{Power System Test Models}
	The modular nature of the EMS platform, from its core functionalities to the web interface, is also reflected in its ability to incorporate many power system models. As part of our work on cyber-attacks and countermeasures, we have studied several different grid models varying in size and complexity. At the moment, the platform includes three systems:
	\begin{enumerate}
	    \item Cascadia is a small-scale synthetic system on the footprint of the grid of Washington state. It is made up of 179 buses, 72 loads, and 37 generators.
	    \item the Polish system available from Matpower \cite{matpower}. This system has 2383 buses, 1822 loads, and 327 generators.
	    \item the synthetic Texas system \cite{Birchfield}; it has 2000 buses, 1125 loads, and 544 generators.
	\end{enumerate}
	New systems can be imported into the EMS platform via an automatic conversion and import function from PSS/E's .raw file format. While this process is fully automated, some manual tuning of the imported system might be required depending on the quality of the original model. Commonly, some tuning will be needed in order to obtain a base case scenario which is feasible under the RTCA and SCED constraints. This might entail modifying generator commitment schedules or generator set-points; in some cases, line limits and load values need to be verified to ensure N-1 feasibility.


\section{\label{sec:Attack}Load Redistribution Attacks}
A false measurement vector $\bar{z}$ created with state attack vector $u$, 
\begin{equation}
\bar{z} = h(x+u)+e, \label{eq:unobservableMeas}
\end{equation}
is \textit{unobservable} to the conventional bad data detector (BDD) embedded within SE, because it is not distinguishable from the true measurements if the true states were $(x+u)$. Given a fixed generation dispatch, unobservable FDI attacks make it appear as the loads are redistributed among load buses, while the total net load remains unchanged. The redistributed loads will cause the system to incorrectly re-dispatch the generators, resulting in physical and/or economic consequences.

Typically, an attacker aims to maximize the attack consequences given its limited resources (\textit{e.g.,} number of measurements controlled). Attacker-defender bi-level linear programs (ADBLPs) can be used to find the worst-case attacks. A general ADBLP is given by 
\begin{subequations}\label{general}
	\begin{flalign}
	\underset{u}{\text{minimize}} \hspace{0.17cm} & c_1^Tu+d_1^Tv^* \label{eq:generalLv1Obj}\\
	\notag \text{subject to} & \hspace{0.08cm}\\
	& A_1u \ge b_1 \label{eq:generalLv1Con}\\
	& v^*=\text{arg}\{\underset{v}{\text{min}} \hspace{0.17cm} d_2^Tv\} \label{eq:generalLv2Obj}\\
	\notag &\text{subject to} \hspace{0.12cm} \hspace{0.08cm}\\	  
	& \hspace{1cm}A_2u+A_3v \ge b_2  \hspace{0.97cm}  \label{eq:generalLv2Con}
	\end{flalign}
\end{subequations}
where $u$ is the attack vector, $v$ is the system decision variable, which can include generation, reserve, load shedding, etc. The attacker's objective \eqref{eq:generalLv1Obj} can be any physical/economic consequence, \textit{e.g.,} maximize the physical power flow on a transmission line, or maximize the operating cost. The attacker's limitations \eqref{eq:generalLv1Con} typically involve its resources (\textit{e.g.,} number of measurements controlled), as well as the load shift resulting from the attack. Here load shift is characterizing the detectability of the attacks, as large load shift may easily trigger system alarm. The defender's problem \eqref{eq:generalLv2Obj}--\eqref{eq:generalLv2Con} models the system response under attack. 

\begin{figure}
		\centering{}\includegraphics[trim=0 0.2cm 0 0.2cm, scale=0.55]{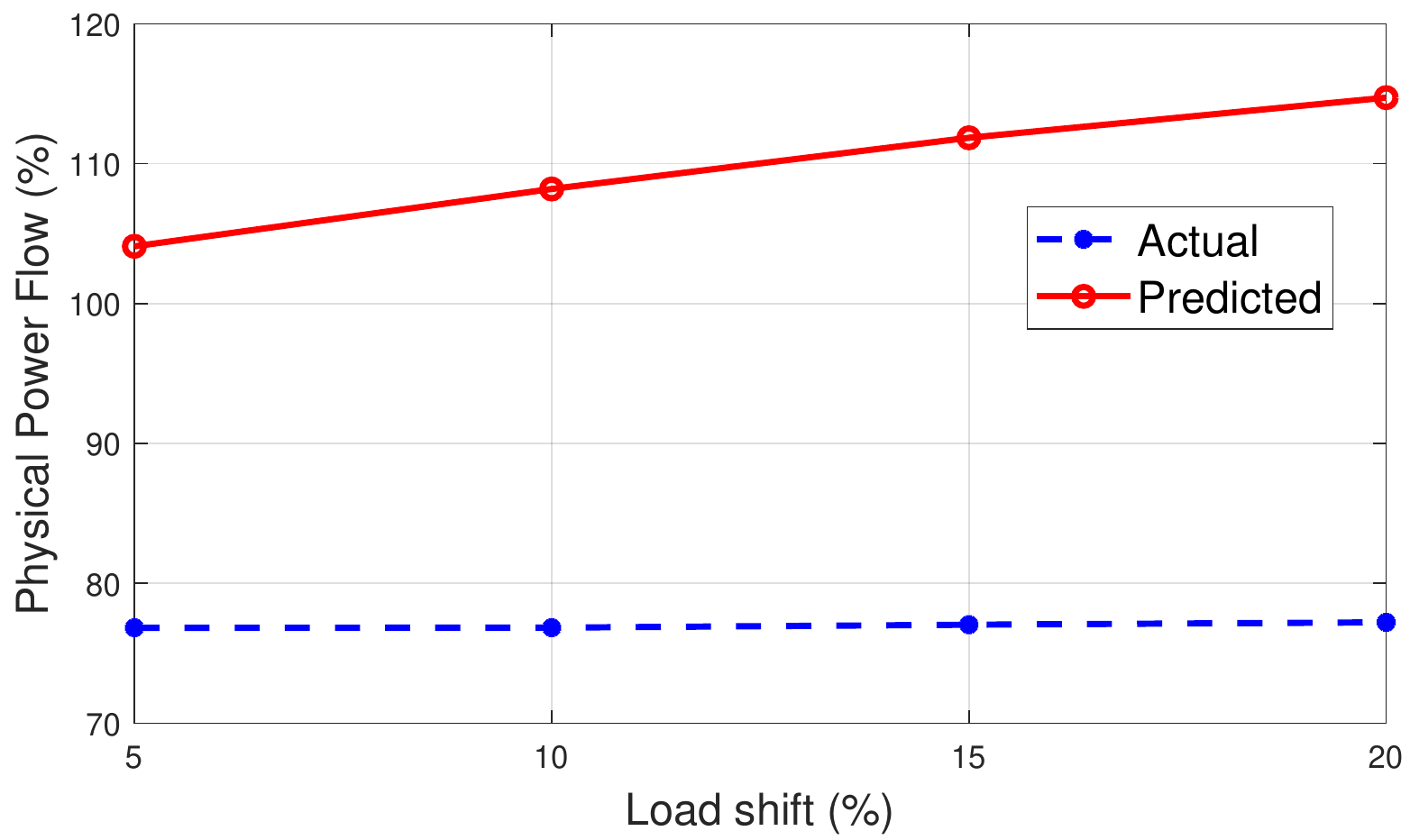}\protect\protect\caption{Consequence of attacks designed considering DCOPF on $N-1$ reliable synthetic Texas system. \label{fig:OAresult}}
	\end{figure}

Prior work \cite{Jzhang2016,Liu2017,Liang2015,Yuan11,Yuan2012,Jia2014} considers only DCOPF as the system response to show the system vulnerabilities to FDI attacks. However, we have found that if the attacker merely solves an ADBLP considering DCOPF as the system response while the system is actually operating using RTCA and SCED as outlined in Fig. \ref{fig:EMS}, the attacker will not accurately predict the system response, and hence, the attacks will not cause the expected consequences. Fig. \ref{fig:OAresult} \cite{Chu2020} illustrates an example based on our experiments on the synthetic Texas system. The attacker solves an attack design ADBLP assuming the system operates using DCOPF, and creates false measurements to launch an attack, aiming to maximize the physical power flow on a target branch. The system performs RTCA and SCED to find the optimal generation dispatch, which yields the actual physical power flows. From this figure, it can be noted that the attacker predicted power flows exceed the rating of the branch for every load shift, but the actual flows do not. 

\begin{figure}
		\centering{}\includegraphics[trim=0 0.3cm 0 0.3cm, scale=0.35]{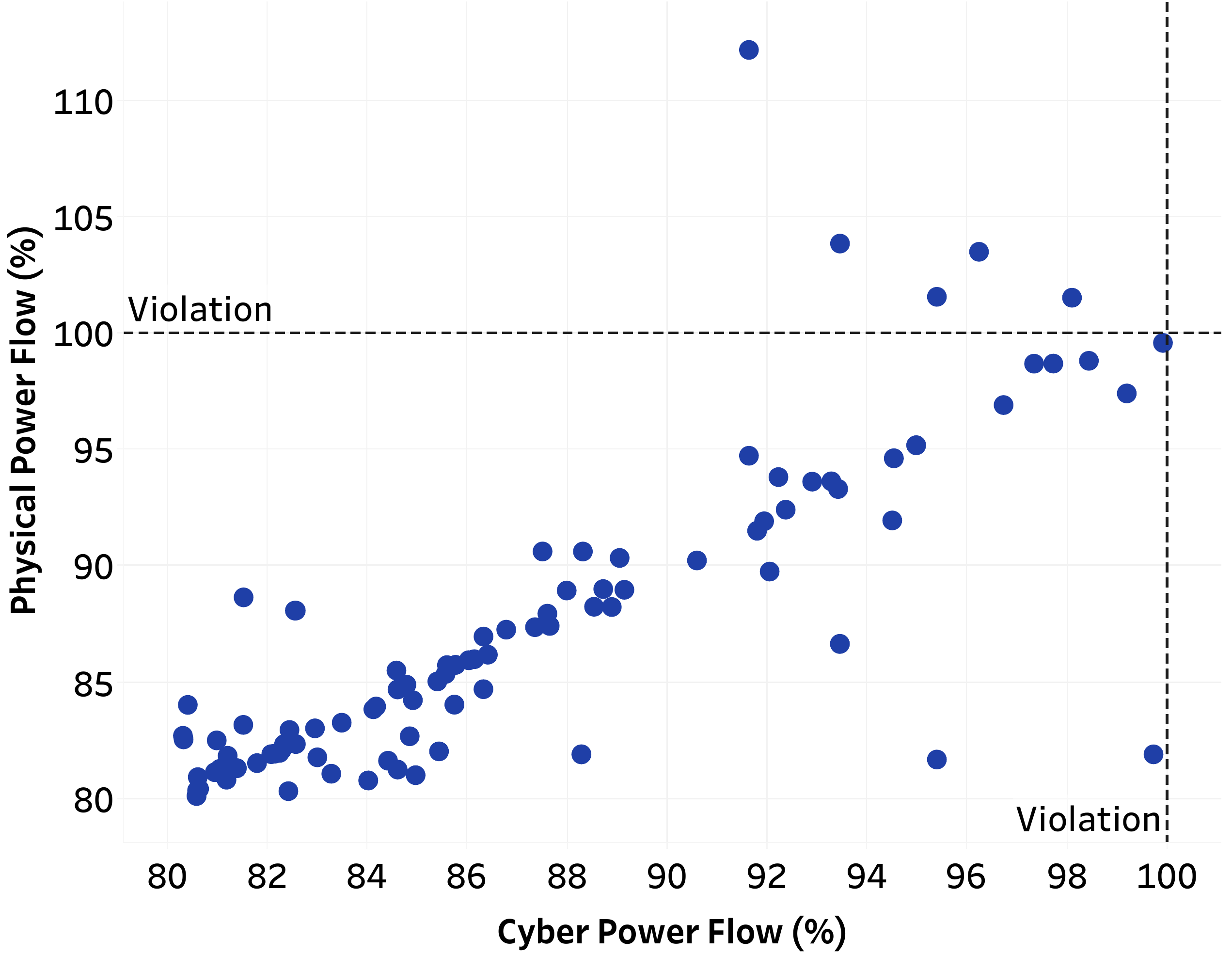}\protect\protect\caption{Cyber (what the operator sees) and physical (what is actually happening in the system) RTCA results after re-dispatch. \label{fig:scatter}}
	\end{figure}

Accurately predicting the system response requires the attacker to gain knowledge of the detailed design of RTCA and SCED used by the system. Although it is not impossible to have this level of knowledge, it is extremely difficult to have such strong attackers in practice. However, those attackers can cause the worst-case consequences on EMSs by modeling SCED as the system response in the attack design ADBLP. In \cite{Chu2020}, such an ADBLP is formulated for the synthetic Texas system and is solved using a modified Benders' decomposition algorithm introduced in \cite{Chu2020SmartGridComm}. The objective of the attacker is to maximize the physical power flow on a target branch, either in the base case or under contingencies. It is shown that no base case overflows can be caused by the attacks due to the conservatism provided by $N-1$ reliability, but post-contingency overflows can still occur, making the system no longer $N-1$ reliable. Fig. \ref{fig:scatter} \cite{Chu2020} compares the physical and cyber RTCA results after the re-dispatch resulting from an attack designed by considering SCED as the system response, aiming to maximize the post-contingency physical power flow on a target branch. The system operators observe the cyber post-contingency power flows on the x-axis, while the post-contingency physical power flows are shown on the y-axis. It can be seen that the attack successfully spoofs the operator by making the system appear to be in a secure state, while in reality, five post-contingency violations are caused by the attack with a maximum post-contingency overflow of 112.2\% on the target branch.
	
	\section{\label{sec:Countermeasure}Data-driven Countermeasures to LR Attacks}
As explained in section \ref{sec:Attack}, FDI attacks against SE effectively result in the system operator estimating slightly modified system loads. For this reason, this type of attacks falls in the class of load redistribution (LR) attacks. 
This observation is the basis for an improved class of detectors that are able to analyse the estimated load values and determine if they are the result of an FDI attack.
This section describes the design of an algorithm that, leveraging the wealth of historical data that is already available to system operators, learns patterns in normative data and checks that the measured loads follow these patterns. 
The basic concept was first explored in \cite{Pinceti2018}, where multiple machine learning techniques were tested and compared. Following the promising results from that work, a nearest neighbor based algorithm has been fully developed to scale to very large systems and attempt to localize the maliciously modified loads \cite{PincetiJournal}. As the culmination of our research, the detector has been integrated in the EMS software platform to showcase its capability and suitability as a tool for system operators. 

\subsection{\label{sec:DetectionAlgo}Attack Detection Algorithm}
The proposed attack detector is based on the idea that the system loads at any given time follow some underlying, recurring patterns; if an attacker arbitrarily modifies a subset of the loads, these patterns will be violated. Nearest neighbor works by finding the minimum distance between the data-point to be tested (i.e. the set of loads at a given time) and a dataset of historical points which are assumed to be attack free (i.e. the historical measured load values). A thresholding approach can then be used to define a point as normal if its distance to the historical data is small or anomalous in case of large distance.

Let us define the vector containing the load values computed from SE as $\mathbf{p}$ and the historical load vector at time $i$ as $\mathbf{h}_i$. The closest distance $d$ for sample $\mathbf{p}$ is defined as
\begin{equation}
d=\min_{r=[1:n\textsubscript{\textit{h}}]} \| \mathbf{p}-\mathbf{h}_r\|\textsubscript{\textsubscript{2}}\label{distance}.
\end{equation}
where $n\textsubscript{\textit{h}}$ is the total number of historical points. The minimum distance $d$ is compared against a predetermined threshold $\tau$ to label the load profile $\mathbf{p}$ as normal or attacked.

As the number of loads in a system increases, the dimensionality of the load vectors gets larger reducing the efficacy of the algorithm. As highlighted in \cite{PincetiJournal}, when computing the Euclidean distance between high dimensional vectors, the effect of the small subset of attacked loads over the entire system is diminishingly small. To overcome this challenge and make the detector scalable to systems of any size, the algorithm has been modified so that it simultaneously analyzes small groups of neighboring loads identifying any suspicious loads. In this case, the minimum distance for each group of loads is computed; 	
\begin{equation}
d_j=\min_{r=[1:n\textsubscript{\textit{h}}]} \|\mathbf{p}^j-\mathbf{h}^j_r\|\textsubscript{\textsubscript{2}}\label{distancegroup}
\end{equation}
where $\mathbf{p}^j$ is the set of real-time values for the loads in group $g_j$, $\mathbf{h}^j_r$ is the subset of loads belonging to group $g_j$ from the $r^{th}$ historical load vector. The minimum distance is then compared to the threshold $\tau_j$ to determine if the loads in group $g_j$ are normative or anomalous; if $d_j > \tau_j$ an alarm for that group is raised. This process is repeated for every group and if one or more alarms are raised, the load vector $p$ is labeled as anomalous.

\begin{figure}
	\centerline{\includegraphics[scale=0.3]{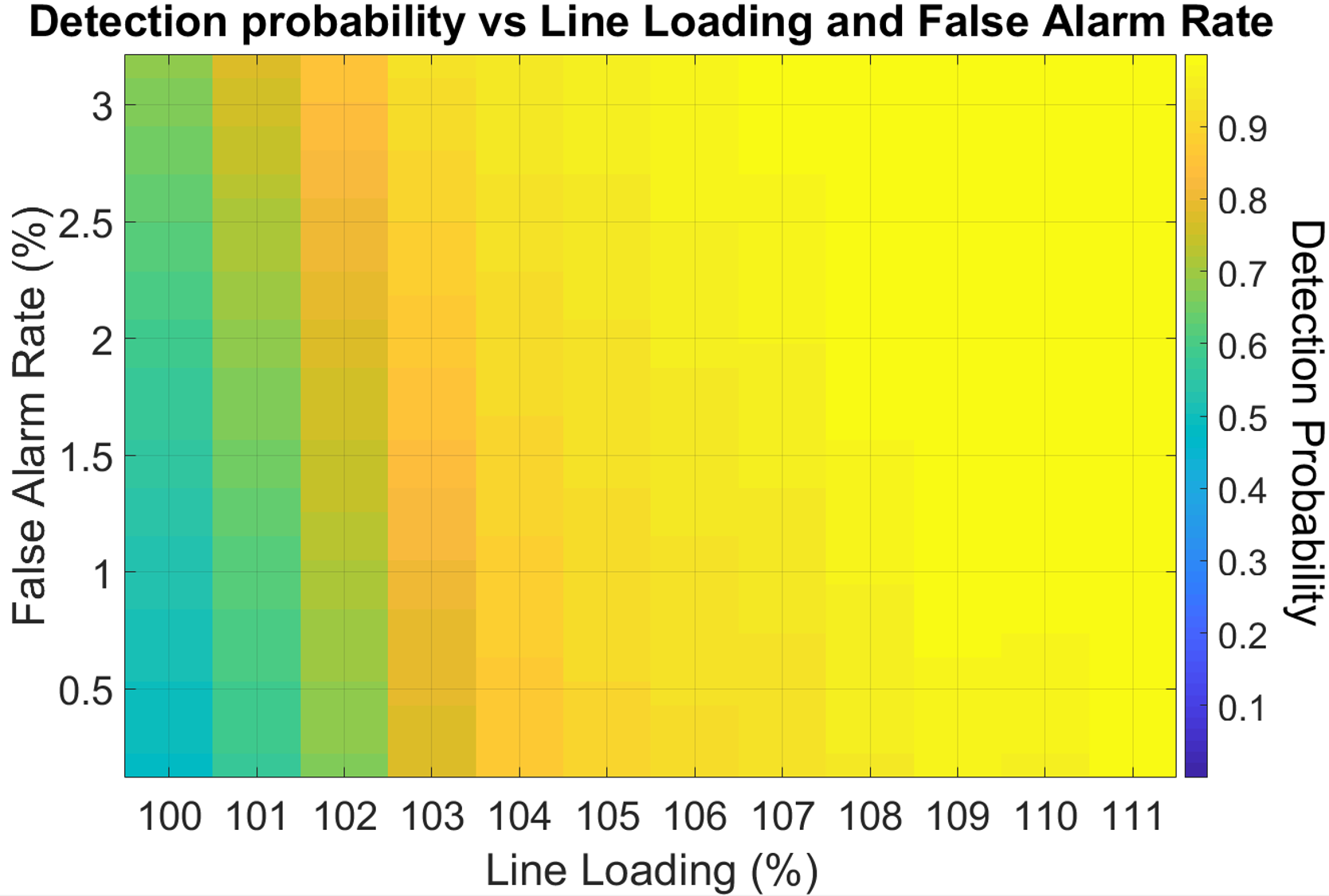}}
	\caption{Detection probability as a function of line overload and false alarm rate.}
	\label{real_DPvsOLandFA} 
\end{figure}

In \cite{PincetiJournal}, the improved algorithm is tested on the synthetic Texas system against a wide range of LR attacks. Figure~\ref{real_DPvsOLandFA} demonstrates the performance of the detector against FDI attacks by showing the detection probability as a function of the percentage line loading on the target line and the estimated false alarm rate. It can be seen that any attack which would result in significant line overloads is detected with almost perfect accuracy. 

\subsection{Further Countermeasure Approaches}
In addition to the nearest neighbor detector that has been implemented in the platform, we have developed and tested several other approaches to attack countermeasures. 

The grouping strategy described in Section~\ref{sec:DetectionAlgo} allows to perform a further risk assessment of the measured system loads and determine how likely each load is to be attacked. After identifying the groups of loads that likely contain values that have been maliciously modified, the loads within each group are analyzed independently. Using a metric called Z-score it is possible to determine how far a load is from its normal behavior, thus characterizing its likelihood of being attacked. This technique and its results are discussed in detail in \cite{PincetiJournal}.

An alternative attack detector, based on support vector models, is presented in \cite{ChuSVM}. Instead of relying solely on the historical data, this detector uses a support vector regression (SVR) model to predict the bus-level loads at the next time step; then, using a support vector machine (SVM), the actual system loads and their forecasted values are analyzed to determine if any difference between them is due to a cyber-attack or normal system behaviors.

Finally, a real-time non-probabilistic approach to detect LR attacks attempting to cause an overflow in smart grids is proposed in \cite{kaviani2020detection}. First, power systems domain insight is leveraged to identify an underlying exploitable structure for the core problem of LR attacks in \cite{kaviani2019identifying}, which enables the prediction of the attackers’ behavior. Then, in the second part of the study, a security index based on the identified structure in \cite{kaviani2019identifying} is developed. This security index can be used in practice, with minimal disruptions in the existing EMSs, to flag LR attacks.

\section{\label{sec:conclusion}Conclusion}
In this paper, the design of an EMS simulation platform is described and it is shown how it allows for the detailed study of cyber-attacks on power systems and their physical consequences.  

Two important and contrasting lessons about the cyber-security of modern energy  management systems can be learned from the results of our simulations and tests. On the one hand, it is shown in Section~\ref{sec:Attack} that the added complexity of systems such as RTCA and SCED provides a further layer of defense against cyber-attacks. Specifically, these tools result in more conservative system operations which in turn yield system states that are harder (if not impossible) to be manipulated by an attacker. Nonetheless, since this added security is not achieved by design as a response to cyber-attacks, but it is simply a byproduct of other power system considerations, it is still possible for a very sophisticated attacker to cause physical consequences and damage to a system. If an attacker gains knowledge of all the control and monitoring tools employed by the operators, it can create attacks that target post-contingency violations. While these are not as immediate of a threat as base case attacks (remember, a physical contingency must happen in order for the unobservable cyber-attack to cause consequences), they still highlight a vulnerability that could be exploited by powerful, well funded attackers. It is for this reason, that we have proposed several attack detection and mitigation schemes.

In general, the EMS platform we built has demonstrated to be a very useful and powerful tool for the analysis of power system operations. Moreover, its modular design structure makes it an ideal platform for the development and implementation of new tools beyond cyber security applications. 

As part of our future work, we are looking at possible improvements and further developments of our attack countermeasures. The detection algorithms can be expanded to include the identification and classification of more classes of cyber-attacks as well as system events and natural phenomena. Moreover, the insight provided by detector based on support vector models can be leveraged for the decision making process following a cyber-attack. In particular, the predicted loads can be used as a replacement for the counterfeit values, practically nullifying the negative effects of the attack.

\section*{Acknowledgment}
This material is based upon work supported by the National Science Foundation under Grants No. CNS-1449080 and OAC-1934766 and the Power System Engineering Research Center (PSERC) under project S-72.

\bibliographystyle{IEEEtran}
\bibliography{CPS_Project}

\end{document}